\documentclass[prb,twocolumn,showpacs]{revtex4}   
\usepackage{graphicx}

\begin{document}

\title{Magnetic transition and orbital degrees of freedom 
in vanadium spinels
}
\author{
Hirokazu Tsunetsugu
}

\affiliation{
Yukawa Institute for Theoretical Physics, Kyoto University, Kyoto
606-8502, Japan
}

\author{
Yukitoshi Motome\footnote{%
Present address: 
RIKEN (The Institute of Physical and Chemical Research), 
Hirosawa 2-1, Wako 351-0198, Japan}
}

\affiliation{
Tokura Spin Superstructure Project, ERATO, 
Japan Science and Technology Corporation, 
c/o National Institute of Advanced Industrial Science and Technology 
Tsukuba Central 4, Tsukuba, Japan
}

\date{May 1, 2003}

\begin{abstract}
We propose a scenario for the two phase transitions 
in $A$V$_2$O$_4$ ($A$=Zn, Mg, Cd), 
based on an effective spin-orbital 
model on the pyrochlore lattice.  
At high temperatures, spin correlations are strongly 
frustrated due to the lattice structure, 
and the transition at $\sim$50 [K] is an orbital order, 
supported by Jahn-Teller lattice distortion.  
This orbital order introduces spatial modulation 
of spin exchange couplings 
depending on the bond direction.  
This partially releases the frustration, and leads  
to a spin order at $\sim$40 [K].  We also study the 
stable spin configuration by taking account of 
third-neighbor exchange couplings and quantum fluctuations.  
The result is consistent with the experimental results.
\end{abstract}

\pacs{75.10.Jm, 75.30.Et, 75.30.Ds, 75.50.Ee}

\maketitle

Pyrochlore lattice is a network of 
corner-sharing tetrahedra shown in Fig.~1, and 
it is a typical 
geometrically frustrated system in three dimensions.  
The vanadium spinel, ZnV$_2$O$_4$, is an 
insulator, and its sublattice of magnetic 
vanadium ions constitutes a pyrochlore lattice. 
It was found that this compound reveals two phase 
transitions at $T_{c1}$=50[K] and $T_{c2}$=40[K].\cite{ZnV2O4}  
An X-ray diffraction 
experiment showed that 
the transition at $T_{c1}$ is 
a structural transition from the high-temperature 
cubic phase to the low-temperature tetragonal phase 
with the lattice constants $a=b>c$.  
The neutron experiment at $T=4.2$[K] showed the presence of 
the antiferromagnetic 
long-range order plotted in Fig.~1,\cite{neutron} 
and the Li-substitute 
material showed an anomaly in its NMR signal 
at $T_{c2}$.\cite{ZnV2O4}  
These indicate that the lower-temperature 
transition in ZnV$_2$O$_4$ is a 
paramagnetic-to-antiferromagnetic transition.  

Theoretical studies have predicted unusual 
properties for spin systems on the pyrochlore lattice. 
In particular, antiferromagnetic classical spin systems 
with only nearest neighbor interactions are 
believed to have no magnetic order at any 
temperature.\cite{Liebmann,Reimers,Moessner}  
It is also believed that the quantum spin systems 
have a spin-singlet ground state and a finite 
energy gap to spin-triplet excitations with 
thermodynamic number of singlet states inside 
the singlet-to-triplet gap.  
Several symmetry breakings are theoretically 
predicted within the spin singlet subspace, 
e.g., dimer/tetramer order, but without 
magnetic long range order.
\cite{Harris,Tsunetsugu,Koga,Berg}  

Therefore, we encounter a difficulty in explaining 
these phase transitions in ZnV$_2$O$_4$, if the material 
is considered as a pure spin $S=1$ system on the pyrochlore lattice, 
and other degrees of freedom are necessary to 
take into account. Yamashita and Ueda studied this 
problem based on a valence-bond-solid (VBS) approach 
and examined the effects of Jahn-Teller distortion.
\cite{Yamashita}
They proposed that the transition at $T_{c1}$ is 
due to the Jahn-Teller effect which lifts the 
degeneracy of the spin-singlet local ground states at each 
tetrahedron unit of the pyrochlore lattice. 
This idea was also applied to classical spin 
systems, and the effects of 
the coupling to the lattice distortion 
were investigated using the point-group argument 
and the Landau theory.
\cite{Tchernyshyov} 
These scenarios are quite appealing, but 
some difficulty still seems to remain.  

\begin{figure}[bt]
\includegraphics[width=6cm]{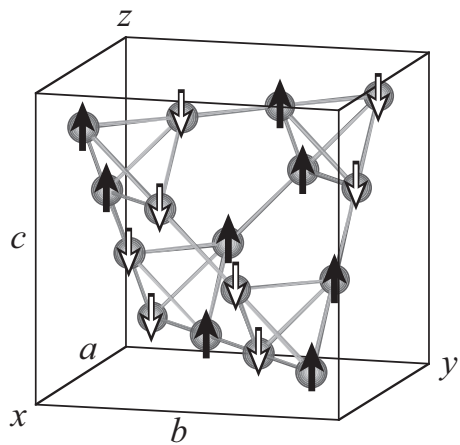}
\caption{
Cubic unit cell of the pyrochlore lattice 
and spin order 
determined by neutron experiments 
at low temperatures.  
}
\label{fig:lattice}
\end{figure}

The problem is that it is difficult to 
explain the magnetic transition at 
the low transition temperature $T_{c2}$ 
by Yamashita-Ueda type scenarios based on 
a quantum-spin picture. 
Like other 
theoretical works, they started from the 
assumption that 
there exists a finite energy gap between 
the spin-singlet ground state and 
spin-triplet excitations, and constructed  
a low-energy effective theory to describe a phase 
transition within the 
spin-singlet sector.  High-energy excitations 
with total spin $S \ne 0$ are already traced 
out from the theoretical framework, and there 
remain no degrees of freedom describing 
the low-temperature magnetic transition.  

Scenarios based on a classical spin picture 
do not have this problem, but it has another 
difficulty to explain the following generic difference 
between the vanadium and chromium spinels.  
The presence of these two transitions is 
common to other vanadium spinels, 
MgV$_2$O$_4$\cite{MgV2O4} and CdV$_2$O$_4$,\cite{CdV2O4} 
with 
the same valence V$^{3+}$.  On the other hand, 
the chromium spinels, ZnCr$_2$O$_4$, 
CdCr$_2$O$_4$, and MgCr$_2$O$_4$ show only one transition at 
12.5[K], 7.8[K], and 12.5[K], 
respectively.\cite{ZnCr2O4,MgCr2O4}  The difference between 
the vanadium spinels and the chromium spinels 
is generic and independent of divalent A-site 
cations.  
It is also unclear whether classical 
approximations are justified at low temperatures 
for the system with the second smallest spin, 
$S=1$.  

In the present study, we will explore another 
scenario to explain the two transitions 
with taking account of the orbital 
degrees of freedom, which exist in 
the vanadium spinels but not in the chromium 
spinels.  The essential difference between 
the vanadium and chromium spinels is 
the number of electrons in magnetic 
ions: V$^{3+}$ ions has $d^2$ configuration and 
Cr$^{3+}$ has $d^3$ configuration.  
Due to the cubic crystal field, 
the $d$-electron orbitals are split into the 
high-energy $e_{g}$ and low-energy $t_{2g}$ multiplets.  
Since in both the spinels the high spin state is realized 
due to large intra-atomic Coulomb interactions, 
two of the 3-fold degenerate $t_{2g}$ orbitals are 
occupied by electrons in the vanadium case, whereas 
all three orbitals are occupied in the chromium case.  
Therefore, each V$^{3+}$ ion has 3-fold orbital 
degeneracy in addition to the triplet spin state $S=1$.  
Rigorously speaking, the crystal field has a small 
trigonal component, which may 
result in a further splitting of the $t_{2g}$ multiplet. 
However, this splitting is compensated by the covalency 
difference of the three states, and the net splitting will 
be small and we neglect this.   

The realistic model Hamiltonian for this system reads 
with the standard notation, 
\begin{eqnarray}
H &=& \sum_{\langle i,j \rangle} \sum_{\alpha \beta} \sum_{\sigma} 
\left[ 
  t_{\alpha \beta} ( {\bf r}_i -{\bf r}_j ) 
  c_{i \alpha \sigma}^\dagger   c_{j \beta  \sigma} 
  + \mbox{H.c.} 
\right] \nonumber\\
&+& {\textstyle {1 \over 2}}
\sum_{i} \sum_{\alpha \beta, \alpha' \beta'} \sum_{\sigma \tau} 
U_{\alpha \beta, \alpha' \beta'} 
c_{i \alpha  \sigma}^\dagger   
c_{i \beta   \tau}  ^\dagger   
c_{i \beta'  \tau}
c_{i \alpha' \sigma} , 
\label{eq:ham}
\end{eqnarray}
where $i,j$ are site index, 
$\sigma, \tau$ are spin index, and 
$\alpha, \beta=$ 1 ($d_{yz}$), 2 ($d_{zx}$), 3 ($d_{xy}$) 
are orbital index.  For the Coulomb interactions, 
we use the standard parameterization, 
$ U_{\alpha \beta, \alpha' \beta'} = 
V \delta_{\alpha \alpha'}\delta_{\beta \beta'} 
+ J ( \delta_{\alpha \beta'}\delta_{\beta \alpha'} 
  + \delta_{\alpha \beta}\delta_{\alpha' \beta'} )$,  
and $U=V+2J$ is the Hubbard interaction in the 
same orbital.  These values were determined from 
the spectroscopy data for the vanadium perovskites 
as, $U \sim $6[eV] and $J \sim$0.7[eV],\cite{Coulomb} 
and it is reasonable to assume a small value 
for the ratio of 
the two in the vanadium spinels, 
$\eta \equiv J/U \sim 0.11$.   
The hopping integrals 
$t_{\alpha \beta} ({\bf r}_i - {\bf r}_j )$ were 
also determined by a parameter fitting of 
the first-principle band calculation data,\cite{hopping} 
and it was found that the one 
for the nearest-neighbor $\sigma$-bond is much 
larger than the others.  
We therefore consider only this kind of hoppings, 
$t_{\sigma}=-0.32$[eV], and set 
the others zero.  
This is the hopping in the case where one of four 
lobes of each $t_{2g}$ orbital is pointing towards 
the other end of the bond.  In this case, the 
number of electrons in each orbital 
is conserved through the hopping processes, 
simplifying the following calculations.  
We neglect the relativistic spin-orbit coupling,  
which is much smaller than the energy scales in 
Eq.~(\ref{eq:ham}).

Since ZnV$_2$O$_4$ is an insulator, we employ the 
strong coupling approach $U \gg t_{\sigma}$, 
and include the hopping 
processes by the second order perturbation.\cite{KK} 
Each vanadium ion V$^{3+}$ has $d^2$ configuration, 
and it is in a high spin state, $S=1$, due to 
large intra-atomic interactions.  
Each vanadium site is represented by 
a spin state $S_{i}^z = 0, \pm 1$ and 
an orbital configuration 
$\{ n_{i \alpha} \}_{\alpha=1}^{3}$, 
which is subject to the local constraint, 
$\sum_{\alpha} n_{i \alpha} =2$, imposed by 
the valence of the vanadium ion.  
Different local states are hybridized by the hopping 
processes, and this is described by an effective 
spin-orbital model of Kugel-Khomskii type 
on the pyrochlore lattice.  
It reads 
\begin{eqnarray}
 &&H_{\rm so} = \sum_{\langle i,j \rangle} 
 \left[ 
   H_{\rm o-AF}^{(ij)} + H_{\rm o-F}^{(ij)}
 \right], 
\label{eq:hamSO}\\
 &&H_{\rm o-AF}^{(ij)} = - K_0 
 (A+B{\bf S}_i \cdot {\bf S}_j) 
\nonumber\\
 && \times 
 \left[ 
    n_{i\alpha(ij)} ( 1 - {n}_{j\alpha(ij)} ) 
  + ( 1 - {n}_{i\alpha(ij)}) n_{j\alpha(ij)} 
 \right], 
\label{eq:ham_oAF}\\
 &&H_{\rm o-F}^{(ij)} = - K_0 C 
 (1-{\bf S}_i \cdot {\bf S}_j) 
 n_{i\alpha(ij)} n_{j\alpha(ij)} , 
\label{eq:ham_oF}
\end{eqnarray}
where $K_0 = t_{\sigma}^2/U >0 $, 
$A=(1-{2 \over 3} \eta)/(1-2\eta)$, 
$B={2 \over 3} \eta/(1-2\eta)$, 
$C=(1+\eta)/(1+2\eta)$, and $\eta = J/U$ 
as defined before.  
$\alpha(ij)$ denotes the orbital in which electron 
hopping is possible between the sites $i$ and $j$. 

The two parts, $H_{\rm o-AF}^{(ij)}$ and $H_{\rm o-F}^{(ij)}$, 
represent the orbital ``antiferro'' and ``ferro''
interactions, respectively.  The sign of 
the spin exchange term in these two indicates that 
an antiferro-orbital configuration favors 
a ferromagnetic spin state, while a ferro-orbital configuration 
favors an antiferromagnetic one, as is usual for the 
Kugel-Khomskii effective model.  
It is noted that the orbital parts contain 
only density-density interactions as 
a consequence of the above-mentioned character 
of the hopping integrals.  This means that the orbital
interaction has a large anisotropy and the 
anisotropy axis depends on the bond direction of 
two neighboring sites, in contrast to 
the spin space with full rotational symmetry. 

We now give a simple argument to discuss the character 
of orbital and spin fluctuations at high temperatures 
where no symmetry breaking takes place.  
To discuss spin fluctuations, we replace orbital 
density operators by their mean value, 
$n_{i\alpha} \rightarrow \langle 
n_{i\alpha}\rangle = {2 \over 3}$, 
and the result is a simple Heisenberg model 
on the pyrochlore lattice with 
the nearest-neighbor coupling 
$J_{s} = {4 \over 9}K_0 (C-B)$ that is 
antiferromagnetic for the realistic value 
of $\eta \sim 0.11$. 
This is a strongly frustrated system and 
the enhancement of spin correlations 
with decreasing temperature will be 
quite small.  The situation in the orbital part 
is different due to its anisotropy.  
This time, we replace the spin operators by their 
mean value, ${\bf S}_i \rightarrow 
\langle {\bf S}_i\rangle = {\bf 0}$, and obtain
the following effective orbital Hamiltonian, 
\begin{eqnarray}
 H_{\rm orb} &=& 
  K_0 \sum_{\langle i,j \rangle}
 \left[ 
 (2A-C) 
  n_{i\alpha(ij)} n_{j\alpha(ij)} 
 \right.
\nonumber\\
  &&  \left. 
  -A ( n_{i\alpha(ij)} + n_{j\alpha(ij)} )
 \right] , 
\label{eq:ham_orb}
\end{eqnarray}
but the linear terms in $n_{i\alpha}$ become
a constant after taking the summation 
over bonds because of the local constraint.  
Since $2A -C \sim 1$ for small 
$\eta$, the orbital interactions are 
``antiferromagnetic'', but 
the number of local orbital states 
is three not two, which corresponds to 
a 3-state Potts model (or equivalently 
a 3-state clock model).
More importantly, the anisotropy axis 
varies from bond to bond depending on its 
direction.  

We now consider the stable orbital configuration 
of $H_{\rm orb}$.  As for a tetrahedron unit, 
there are essentially two different types 
of stable configurations shown in Fig.~2.  
In the first type (a), 
two ferro-orbital bonds do not touch each 
other, while they touch at one site in the 
second type (b), and 
these two types have the same energy, $-(4A+2C) K_0$.  
Aside from the examples shown in Fig.~2, 
there are 2 and 3 other equivalent configurations 
belonging to the first and second type, 
respectively.  

\begin{figure}[tb]
\includegraphics[width=6.5cm]{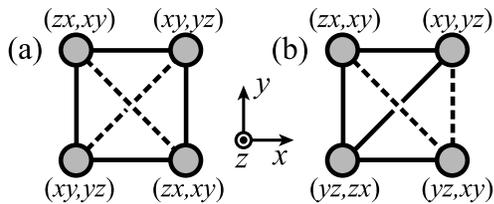}
\caption{
Mean-field ground states of the effective 
orbital model Eq.~(\ref{eq:ham_orb}).  
Configuration in a tetrahedron unit cell 
is shown.  
Labels, $xy$ et al., indicate two occupied 
orbitals at each site.  
Solid (dash) lines denote antiferro-orbital 
(ferro-orbital) bonds, which have 
ferromagnetic (antiferromagnetic) spin 
couplings. 
}
\label{fig:orbital}
\end{figure}

The degeneracy in energy between the two types of configurations 
is lifted if the Jahn-Teller coupling is taken into account. 
Below the higher transition temperature $T_{\rm c1}$, 
the system is compressed along the $c$ axis.  
Considering the corresponding shift of oxygen atoms along 
the $c$ axis, the energy level of $d_{xy}$-orbital 
is pushed down relative to the other orbitals, 
and we write the level separation as $\Delta_{\rm JT}$.  
The energy correction is then 
$-{4 \over 3} \Delta_{\rm JT}$ and 
$-{1 \over 3} \Delta_{\rm JT}$ for the two orbitals 
configurations shown in Fig.~2 (a) and (b), respectively.  
The energy difference is due to the difference 
in the number of occupied $d_{xy}$ orbitals.  
Therefore, it is natural 
to understand that the higher-temperature 
transition is an orbital ordering of the type (a).  

This orbital order introduces a spatial modulation 
of spin exchange couplings, and leads to the reduction 
of the spin frustration as a consequence.  
As seen from Eqs.~(\ref{eq:hamSO})-(\ref{eq:ham_oF}), 
a ferro-orbital bond has an 
antiferromagnetic spin coupling, ($J_{\rm 1A} = K_0 C$), 
while an antiferro-orbital bond has a ferromagnetic 
coupling ($J_{\rm 1F} = -K_0 B$).  The spatial pattern of 
spin exchange couplings is shown in Fig.~3(a).  
Since the antiferromagnetic couplings  
are much stronger than the ferromagnetic ones, 
$J_{\rm 1A} \gg | J_{\rm 1F} |$ for the realistic 
value of $\eta$, antiferromagnetic 
spin alignment is stabilized in each chain 
in the $xy$-planes, and the entire spin configuration 
will be built up by stacking of antiferromagnetic 
chains.   

\begin{figure}[tb]
\includegraphics[width=6.5cm]{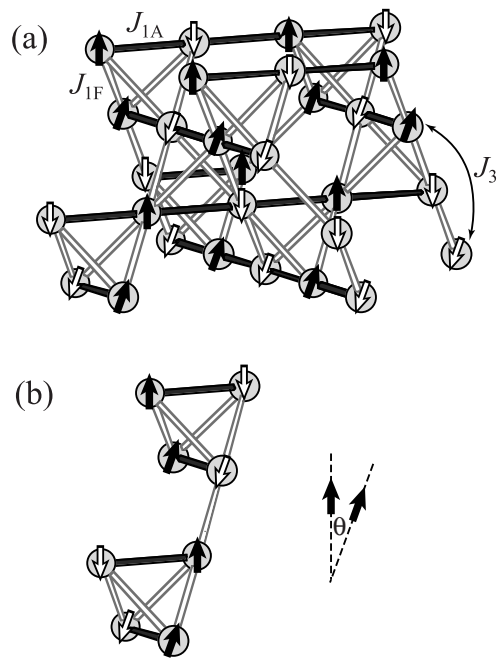}
\caption{
(a) Spin exchange couplings in the orbital ordered 
state.  Strong antiferromagnetic couplings and 
weak ferromagnetic couplings are shown by 
black and white bonds, respectively.  
$J_3$ is the third-neighbor 
interaction. 
(b) Magnetic unit cell of the ${\rm q}=(0,0,2\pi /c)$ 
state.  
}
\label{fig:chain}
\end{figure}

Interactions between the antiferromagnetic chains are 
ferromagnetic ones, $J_{\rm 1F}$, but 
frustrated as shown in Fig.~3(a).  The frustration is 
due to the structure of the pyrochlore lattice 
and also antiferromagnetic spin correlations in each 
chain, but the sign of interchain couplings $J_{\rm 1F}$ 
is not essential.  Therefore, the relative angle of 
spins between different chains remains undetermined 
in the mean-field level approximation for the spin-orbital 
model $H_{\rm so}$, and it is determined by some other 
mechanisms.  

The first mechanism to consider 
is longer range exchange interactions, 
in particular, the third neighbor interactions, $J_3$.  
Fitting of the band calculation results 
predicted that the $\sigma$-bond of third-neighbor 
pairs has a larger amplitude than for second-neighbor 
pairs,\cite{hopping} and, more importantly, 
exchange couplings are frustrated between second-neighbor 
pairs.   
The third-neighbor exchange coupling has a large amplitude,  
and it is also antiferromagnetic.  Therefore, 
this implies an order with 
${\bf q}=(0,0,2\pi/c)$.  
Two tetrahedron unit cells in two neighboring  
$xy$-planes have spin configurations opposite to each other. 
However, the relative angle, $\theta$, between the spins in the bottom 
plane of each tetrahedron and those for
the top plane is not yet determined.  

The second mechanism is fluctuations of spins, and 
we here determine the relative angle $\theta$ 
which minimizes the 
the zero-point energy of quantum fluctuations.\cite{Shender}  
We use the standard 
spin wave approach starting with the magnetic unit 
cell containing 8 sites shown in Fig.~3(b).  
By means of the Bogoliubov 
transformation, the energy dispersion of magnons 
is calculated, $\omega_{{\bf k}\gamma}$, where 
$\gamma$ labels the magnon branch, and the 
zero-point energy is obtained by 
$ E_{\rm 0pt} (\theta ) = \Omega^{-1} 
\sum_{{\bf k}\gamma} \omega_{{\bf k}\gamma} $, 
where $\Omega$ is the number of sites 
and $\hbar$=1 in our units.  
Calculations have been performed for various 
values of $J_{\rm 1F}/J_{\rm 1A}$, and 
typical results are shown in Fig.~4. 
Here we set 
$J_3=0$ for simplicity, since this term is not 
essential for determining $\theta$.  
The zero-point energy is always minimum at 
$\theta =0, \pi$, which means that collinear 
order is stablest, and these two are equivalent, 
since they are related to each other 
by the mirror symmetry with respect to the $x=y$ plane.  
This spin order agrees with the experimental result 
shown in Fig.~1.  

\begin{figure}[tb]
\includegraphics[width=6cm]{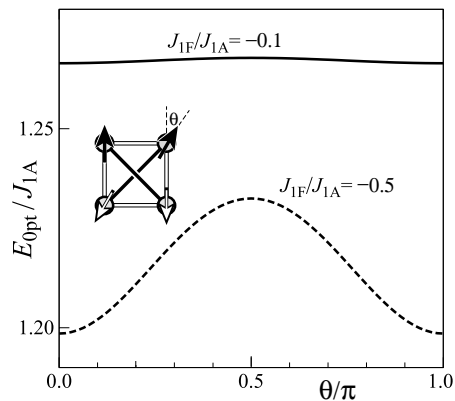}
\caption{
Zero-point energy of quantum fluctuations 
as a function of $\theta$, the angle of 
local staggered moments between the 
neighboring chains.  
}
\label{fig:eng0}
\end{figure}

In this paper, we have proposed a scenario for 
the two phase transitions in the vanadium spinels 
$A$V$_2$O$_4$ ($A$=Zn, Mg, Cd).  
In our scenario, the high-temperature transition at 50[K] is 
an orbital order assisted by the Jahn-Teller 
distortion.  This orbital order introduces 
spatial modulation of spin exchange couplings.  
The geometrical spin frustration is consequently 
relaxed, and this 
induces the low-temperature magnetic transition 
at 40[K].  
If there is no orbital order, the system remains 
subject to strong frustration, and we believe that 
this is the case for $A$Cr$_2$O$_4$ ($A$=Zn, Mg, Cd).
In those compounds, since Cr$^{3+}$ ion has the 
$(3d)^3$ electron configuration, 
it has a spin $S=3/2$ and no orbital degrees of freedom.
Those compounds show only one transition at 
$\sim$10[K], and we believe that this is driven 
by a coupling to lattice distortion as proposed 
in Ref.~\onlinecite{Tchernyshyov} .  A similar transition can occur 
at a low temperature in the vanadium spinels in principle. 
However, since the orbital order and subsequent magnetic 
transition occur at higher temperatures, 
this type of spin Jahn-Teller transition does not 
occur in the vanadium spinels in reality.

This work is supported by a Grant-in-Aid from the Ministry of 
Education, Science, Sports, and Culture. 
Parts of the numerical computations were performed on supercomputers 
at the Institute of the Solid State Physics, 
University of Tokyo, 
and Yukawa Institute Computer Facility, Kyoto University.


\end{document}